\documentclass[fleqn,usenatbib]{mnras}
\usepackage{multicol}
\usepackage[T1]{fontenc}
\usepackage{ae,aecompl}
\usepackage{graphicx}	% Including figure files
\usepackage{amsmath}	% Advanced maths commands
\usepackage{amssymb}	% Extra maths symbols
\usepackage{multirow}
\title[Cosmology in a locally scale invariant gravity]{Cosmology in a locally scale invariant gravity}

\author[Meir Shimon]{Meir Shimon$^{1}$\thanks{E-mail: meirs@tauex.tau.ac.il}
\\ 
$^{1}$ School of Physics and Astronomy, Tel Aviv University, Tel Aviv 69978, Israel\\
}

\date{Accepted XXX. Received YYY; in original form ZZZ}

\pubyear{2022}

\begin{document}

\label{firstpage}
\pagerange{\pageref{firstpage}--\pageref{lastpage}}
\maketitle 
\begin{abstract}
A `bouncing' cosmological model is proposed in the context 
of a Weyl-invariant scalar-tensor (WIST) theory of 
gravity. In addition to being Weyl-invariant the theory is U(1)-symmetric 
and has a conserved global charge. The entire cosmic background 
evolution is accounted for by a complex scalar field 
that evolves in the {\it static} `comoving' frame. 
Its (dimensional) modulus $\chi$ regulates the 
dynamics of masses and the apparent space expansion. 
Cosmological redshift is essentially due to the cosmic 
evolution of the Rydberg constant in the comoving frame. 
The temporal evolution of $\chi$ is analogous to 
that of a point particle in the presence of a central 
potential $V(\chi)$. The scalar 
field sources the spacetime curvature; as such it 
can account for the (cosmological) Dark Sector.
An interplay between the energy density of radiation 
and that of the kinetic energy associated with the 
phase $\alpha$ of the scalar field (which are of 
opposite signs) results in a {\it classical} 
non-singular {\it stable} and nearly-symmetric 
bouncing dynamics deep in the radiation-dominated era. 
This encompasses the observed redshifting era which preceded 
by a `bounce' that follows a blushifting era. 
The model is essentially free of the horizon or 
flatness problems. Big Bang nucleosynthesis sets a 
{\it lower} 1-10 MeV bound on the typical energy scale 
at the `bounce'.
\end{abstract}


\section{Introduction}
General relativity (GR), the backbone of the standard 
cosmological model, has successfully passed numerous 
tests within our solar system. However, it is not 
comparably successful on larger, galactic and supergalactic 
scales, unless cold dark matter (CDM) and dark energy (DE) are 
introduced into the cosmic energy budget. The latter are 
foreign to the standard model (SM) of particle 
physics and only appear 
in our cosmological model as nearly perfect fluids 
with characteristic equations of state (EOS).
   
In addition, this classical field theory of gravitation 
is genuinely plagued by singularities. A few singularity 
theorems imply that curvature is inevitably 
singular unless certain plausible `energy conditions' are 
violated. For the latter to take place some form 
of exotic matter is needed. 
Curvature or energy density singularities are 
encountered either at the centers of black 
holes or at the Big Bang, even in the presence 
of a very early inflationary phase in the latter case 
\citeauthor{Borde et al.2003} (\citeyear{Borde et al.2003}). 
It is widely hoped that a 
would-be quantum theory of gravity will ameliorate these 
unwelcome singularities, thereby constituting a major 
thrust behind the quest for a quantum theory of gravity.   

The largest physical scales, over which the short-range 
nuclear interactions and the highly screened 
electromagnetic interaction are irrelevant, are an ideal 
testbed for alternative theories of gravitation, and indeed 
a few major persistent anomalies of $\Lambda$CDM 
may possibly indicate that GR requires modifications on 
cosmological scales.

The GR-based $\Lambda$CDM, with an early inflationary 
phase and a dominant dark sector [the latter contains 
CDM and DE components that determine the background 
evolution, large scale structure (LSS) 
formation history, and gravitational 
potential wells on galactic and supergalactic scales], 
has clearly proved to be a very successful paradigm that 
provides a compelling interpretation of essentially 
all currently available cosmic microwave background 
(CMB) and LSS observational data, 
as well as of light element abundances based on Big Bang 
nucleosynthesis (BBN).
It is remarkable that $\Lambda$CDM provides a very good 
fit to this extensive observational data, that sample 
a wide range of phenomena over a vast dynamical range, 
using less than a dozen free parameters. 

However, $\Lambda$CDM also lacks in a few ways.
A major drawback is that the essence 
of DE and CDM  remains elusive. 
Another problem is that, the currently leading inflationary 
scenario, `eternal inflation', seems to lack 
predictive power as it is most naturally 
realized in the multiverse. 
In addition, conceptually, the existence of the Big Bang, 
which essentially signals the breakdown of 
the underlying theory of gravitation, is also 
a major problem.

Moreover, a few mild to strong inconsistencies 
between various datasets comparison to 
$\Lambda$CDM have been found, 
e.g. \citeauthor{Addison et al.2016} (\citeyear{Addison et al.2016}).
These include the existence of a persistent relative deficit in power 
of density perturbations on superhorizon scales, e.g. 
\citeauthor{Hinshaw et al.1996} (\citeyear{Hinshaw et al.1996}), 
\citeauthor{Copi et al.2009} (\citeyear{Copi et al.2009}),
\citeauthor{Schwarz et al.2016} (\citeyear{Schwarz et al.2016}), 
an anomalously large weak lensing of the 
CMB anisotropy by the intervening large scale 
structure between the present and the last scattering 
surface [\citeauthor{Planck Collaboration et al.2020} 
(\citeyear{Planck Collaboration et al.2020})], 
a statistically significant `Hubble tension' 
between local and high-z inferences of the local 
expansion rate, e.g. \citeauthor{Freedman2017} 
(\citeyear{Freedman2017}), 
\citeauthor{Riess et al.2018} (\citeyear{Riess et al.2018}), 
\citeauthor{Riess et al.2019} (\citeyear{Riess et al.2019}),
\citeauthor{Wong et al.2020} (\citeyear{Wong et al.2020}), and others. 

The main objective of the present work is to demonstrate 
the viability of an alternative, classical, non-singular 
`bouncing' cosmological model within a physical framework 
based on a globally U(1)-symmetric and locally 
Weyl-invariant scalar-tensor (WIST) theory of gravity, 
while the SM of particle physics is left unchanged. 
Unlike in quantum gravity-inspired 
bouncing models classical bounces 
may take place at typical energies that are well within 
the range of well-established physics.

The terminology that will be used in this work is 
somewhat different from the one used in the standard cosmological model. 
In the latter, the observed redshift on cosmological scales is synonymous 
to space expansion, and in bouncing cosmological models it is actually meant that 
space contraction itself momentarily halts at the `bounce' followed by 
space expansion. Since in the proposed model space is static and cosmic 
evolution is regulated by the temporal evolution of mass we will employ 
the more appropriate notions of `turning point', `blueshifting'- and 
`redshifting-phase' instead of the commonly used parlance 
of `bounce', `contraction' and `expansion', respectively. 

Throughout, we adopt a mostly-positive signature for 
the spacetime metric $(-1,1,1,1)$. Our units convention 
is $\hbar=c=1$. We outline the theoretical 
approach adopted in the present work in section 2, 
and the cosmological model is presented in section 3. 
The main results are summarized in section 4. 
Stability analysis at and around the `bounce' is 
carried out in Appendix A.

\section{THEORETICAL FRAMEWORK}

The gravitational sector of the fundamental interactions is 
described by the following WIST action which is in 
addition globally U(1) symmetric 
[\citeauthor{Shimon2021b} (\citeyear{Shimon2021b})]
\begin{eqnarray}
\label{eq1}
\mathcal{I}_{WIST}&=&\mathcal{I}_{gr}+\mathcal{I}_{S},
\end{eqnarray}
where
\begin{eqnarray}
\label{eq2}
\mathcal{I}_{gr}&\equiv&\xi\int\left(|\phi|^{2}R
+6\phi^{*}_{\mu}\phi^{\mu}\right)\sqrt{-g}d^{4}x\\
\label{eq3}
\mathcal{I}_{S}&\equiv&\int V(|\phi|;\{\psi\})\sqrt{-g}d^{4}x, 
\end{eqnarray}
are the `free gravitational' and `source' actions, 
respectively, $\xi$ is a dimensionless parameter, 
and $\phi\equiv\chi e^{i\alpha}$.

Here and throughout, Greek 
indices run through spacetime coordinates, 
$f_{\mu}\equiv\frac{\partial f}{\partial x^{\mu}}$ 
for any scalar function $f$, and summation 
convention applies.
The curvature scalar $R$ is calculated from the 
spacetime metric $g_{\mu\nu}$ and its 
first and second derivatives in the usual way.

The potential $V(\chi;\{\psi\})$ depends on both 
the modulus $\chi$ of the scalar field and 
$\{\psi\}$. The latter collectively denotes all 
fields other than $\phi$, including $g_{\mu\nu}$. 
Aside from being non-negative, the potential $V$ is only 
required to satisfy the relevant field equations, and may 
in general be unrelated to ordinary matter.
Since $V$ is independent of the phase $\alpha$, the latter 
is a `cyclical coordinate' in field space that 
appears in Eq. \eqref{eq1} only via its derivative coupling. 
Consequently, Eq. \eqref{eq1} has a global U(1) symmetry 
and a conserved global charge.

By construction, Eqs. \eqref{eq2} and \eqref{eq3} are (Weyl-) 
invariant under $g_{\mu\nu}\rightarrow\Omega^{2}(x)g_{\mu\nu}$, 
$|\phi|\rightarrow|\phi|\Omega^{-1}(x)$ and 
$V\rightarrow V\Omega^{-4}$. 
For the kinetic term to have the canonical 
form $\phi^{*}_{\mu}\phi^{\mu}$ (up to a sign difference) 
$\xi$ must be 1/6. Hereafter, the WIST theory considered in 
the present work consists of Eqs. \eqref{eq1}-\eqref{eq3} with $\xi=1/6$. 

According to Eq. \eqref{eq1}, spacetime curvature is 
sourced by the kinetic and potential terms that 
appear in Eqs. \eqref{eq2} and \eqref{eq3}, respectively. 
This is not a radically different situation than in 
the GR-based $\Lambda$CDM model where gravitation 
is not exclusively determined by $\mathcal{L}_{SM}$, 
the matter lagrangian density of the SM of particle 
physics; in order for $\Lambda$CDM to provide a 
reasonably good fit to observational data 
$\mathcal{L}_{SM}$ is amended by 
CDM and DE which together comprise up to $\sim 95\%$ 
of the cosmic energy budget at present. However, 
a notable conceptual difference is that in the framework 
described by Eqs. \eqref{eq1}-\eqref{eq3} a single scalar field, 
$\chi$, is responsible to the evolution of the Planck mass, 
active gravitational masses, and potentially also 
to the existence of what is normally interpreted as 
CDM and DE on cosmological scales. In contrast, in 
$\Lambda$CDM the Planck mass is fixed and the evolution is 
determined by space expansion, which in turn is driven the 
by energy budget of which CDM and DE are key building blocks. 
The former is believed to be in the form of some exotic 
beyond-the-SM particles, and the latter is believed to 
be the manifestation of some slow-rolling quintessence field.

The field $\chi$ is the analog 
of $a(t)$, in $\Lambda$CDM, 
but in the latter case the dynamics represents space 
expansion and energy conservation of the individual 
species. The latter are described by effective perfect 
fluids. In other words, the 
total energy density, $\rho$, in the standard 
$\Lambda$CDM model is $\rho=\sum_{i}\rho_{i}$ 
with $\rho_{i}\propto a^{-3(1+w_{i})}$ being the 
energy density of the i'th species which is 
characterized by an EOS, $w_{i}$. 
It is assumed that they do not interact, i.e. 
energy-momentum is separately conserved for 
the individual species, and consequently each 
species evolves independently of the others. 
In contrast, in the WIST-based model, 
the energy-momentum is not the source of curvature, yet 
we obtain analogous evolution to that of $\Lambda$CDM 
simply because it is assumed that the potential, $V$, 
is an analytic {\it polynomial} in $\chi$, which is 
arguably one of the simplest possible shapes that 
a potential could possibly have.

Since all quantities 
that would be normally referred to as `active 
gravitational masses', as well as the `Planck 
mass', are now governed by a single scalar field $\chi$, 
the dimensionless ratios of active gravitational 
masses of any two bodies, as well as ratios of active masses 
to the Planck mass, are fixed to their standard values. 

If the scalar field is fixed $\chi=\sqrt{\frac{3}{8\pi G}}$
and $V(\chi)$ is renamed $\mathcal{L}_{M}$ then  
Eq. \eqref{eq1} reduces to the Einstein-Hilbert (EH) action. 
The specific constant $G$ that appears in the EH action 
guarantees that the resulting gravitational field equations 
reduce to the Poisson equation in the weak field limit 
within, e.g. the solar system, where the `universality' of 
$G$ has been reasonably established. In addition, even if 
we favor the idea that DM exists in the form 
of some exotic, beyond-the-SM particles, 
we still lack Cavendish-like {\it experimental} evidence that 
they `couple' gravitationally via the same `Universal' 
strength $G$ either to each other or to baryons. 

General properties of the theory described by 
Eqs. \eqref{eq1}-\eqref{eq3} are discussed in 
\citeauthor{Shimon2021b} (\citeyear{Shimon2021b}). 
For example, the tensor 
$S_{\mu\nu}\equiv -\frac{2}{\sqrt{-g}}
\frac{\delta(\sqrt{-g}V)}{\delta g^{\mu\nu}}$, 
satisfies
\begin{eqnarray}
\label{eq4}
S^{;\nu}_{\mu\nu}=\phi_{,\mu}\frac{\partial V}{\partial\phi}
+\phi_{,\mu}^{*}\frac{\partial V}{\partial\phi^{*}}, 
\end{eqnarray}
i.e. in general it is not conserved. This is expected 
since the Planck mass 
varies in space and time. In addition, the requirement that 
$\mathcal{I}_{S}$ described by Eq. \eqref{eq3} is WI implies that 
\begin{eqnarray}
\label{eq5}
\phi\frac{\partial V}{\partial\phi}
+\phi^{*}\frac{\partial V}{\partial\phi^{*}}=S, 
\end{eqnarray}
where $S\equiv S_{\mu}^{\mu}$, and assuming that 
only variations of $g_{\mu\nu}$ and 
the scalar field are considered. In analogy with 
the energy-momentum tensor 
$T_{\mu}^{\nu}= -\rho\cdot diag(1,-w,-w,-w)$ of 
a perfect fluid characterized by an EOS $w$, that is 
employed in the standard cosmological model, we have 
$S_{\mu}^{\nu}= -V\cdot diag(1,-w,-w,-w)$. The perfect 
fluid assumption in cosmology is the embodiment of the idea 
that there is no energy and momentum flow in an isotropic 
and homogeneous Universe. The microphysical meaning of 
$w\equiv P/\rho$, where $P$ is the pressure, 
is irrelevant in $\Lambda$CDM.  
In practice, the EOS $w$ is simply a parameter that determines 
the evolution rate of the energy density, and so is the case 
in the model that is proposed here. 

Since $S_{0}^{0}=-V$ it immediately follows 
from Eq. \eqref{eq5} that
\begin{eqnarray}
\label{eq6}
V\propto\chi^{1-3w},
\end{eqnarray}
in the case of an effective single fluid with an EOS $w$, 
\citeauthor{Shimon2021b} (\citeyear{Shimon2021b}).
Consequently, the potential is quartic in the case $w=-1$, 
is independent of $\chi$, i.e. of active masses, 
in the case $w=1/3$, and is linear in masses in 
case of non-relativistic (NR) matter, i.e. the case $w=0$. 
Although there is no fundamental principle 
that requires $V$ to be an analytic function of the 
field $\chi$ we do impose this restriction. 
In particular, we assume that $V$ is an analytic polynomial. This by itself 
corresponds to a constraint on the EOS, $w\leq 1/3$, in case that 
$V$ was a monomial.
As is generically the case, there is no prescription for choosing 
the form of the potential. The latter is designed, subject 
to certain symmetry requirements, to recover (along with 
the kinetic terms) 
the required dynamics of the fields, as determined by 
experiments/observations. Therefore, and following 
the foregoing discussion, the `coefficients' $V_{i}({\psi})$ 
in the potential
\begin{eqnarray}
\label{eq7}
V(|\phi|;\{\psi\})=\sum_{i=0}^{i_{max}}V_{i}(\{\psi\})\chi^{i}, 
\end{eqnarray}
are determined on cosmological scales by the observed 
cosmic evolution. 
The effective EOS associated with the i'th contribution is
\begin{eqnarray}
\label{eq8}
w_{i}\equiv\frac{1-i}{3}. 
\end{eqnarray}
Consequently, the only allowed terms are characterized 
by parameters $w_{i}$ which are integer multiples of 1/3 
subject to the constraint $w_{i}\leq 1/3$ for all i, 
$V$ is a polynomial in $\chi$ that contains only 
non-negative powers of $\chi$ and there is no 
`anisotropy problem'. The latter is a well-known 
problem \citeauthor{Misner1969} (\citeyear{Misner1969}) 
that is generic to bouncing models 
\citeauthor{Levy2017} (\citeyear{Levy2017}).
In practice, the constraint the $w_{i}\leq 1/3$ is implicitly 
assumed to hold in the standard $\Lambda$CDM model as well.
The `mixmaster' model of 
\citeauthor{Misner1969} (\citeyear{Misner1969}) typically 
arises from allowing for different evolutions along three 
principal axes. This would require, in particular, that 
these three functional degrees of freedom are manifested in 
the matter lagrangian density. In the limit of small 
anisotropic evolution, this is represented by an effective 
stiff matter contribution to the energy budget that dominates 
the cosmic energy budget prior to radiation.
This dynamics does not have an analog in the present model 
because the potential $V$ is `protected' against this 
`shearing' dynamics by the assumed U(1) symmetry.

In the same fashion that effective energy density and 
pressure of the various species in $\Lambda$CDM depend 
on a single $a(t)$ function, commensurate with 
isotropic expansion, so does $V$ in Eq. \eqref{eq7} 
depends on $\chi$, and not on $\chi\cos\alpha$ 
and $\chi\sin\alpha$. In other words, the underlying 
U(1) symmetry of Eq. \eqref{eq7} reflects isotropic 
expansion in the $\Lambda$CDM model.

\section{COSMOLOGICAL MODEL}
The modulus of the scalar field, $\chi$, delineates 
essentially the same dynamics that the scale 
factor $a(\eta)$ does in $\Lambda$CDM 
(insofar $\alpha$ is dynamically insignificant). 
However, unlike $a(\eta)$ which is part 
of the Friedmann-Robertson-Walker (FRW) metric, $\chi(\eta)$ is a scalar 
field living on a static spacetime.
If the time variation of $\alpha$ is sufficiently slow, 
the entire observable cosmic evolution, from BBN (taking place 
at typical $\sim $1 MeV energies) onward, is essentially
indistinguishable from that of $\Lambda$CDM, thereby 
retaining its merits. However, the very early 
Universe can be much different, e.g. there is no 
initial singularity in the proposed model and possibly also 
no primordial phase transitions that in $\Lambda$CDM 
are expected to have taken place at energy scales 
of O(100) MeV and O(100) GeV, and possibly 
also at the GUT scale.

\subsection{Evolution of the cosmological background}

The equivalence between the FRW action and Eq. \eqref{eq1}  
in the non-vacuum homogeneous and isotropic case, 
in the case of a {\it real} $\phi$ field, has been 
discussed in, e.g. \citeauthor{Shimon2021b} (\citeyear{Shimon2021b}).
As mentioned above, according to the alternative 
cosmological model 
proposed here, invoking a complex scalar field 
evades the initial singularity problem thereby 
resolving in effect the cosmological 
horizon \citeauthor{Rindler1956} (\citeyear{Rindler1956}), 
\citeauthor{Misner1969} (\citeyear{Misner1969}),
and flatness \citeauthor{Dicke1969} (\citeyear{Dicke1969}) problems. 
In the following we derive the background evolution equations. 

Before focusing on the case of Minkowski spacetime background 
we consider a more general spacetime which is described by the 
static metric $g_{\mu\nu}=diag(-1,\frac{1}{1-Kr^{2}},
r^{2},r^{2}\sin^{2}\theta)$. The latter is the FRW metric 
in the `comoving frame' where the time 
coordinate is conformal time $d\eta=dt/a$, 
and $K$ is the spatial curvature parameter. This spacetime 
trivially satisfies the Cosmological Principle.
The Ricci curvature scalar associated with this metric 
is $R=6K$. Considering $\phi=\chi e^{i\alpha}$ and 
assuming $\xi=1/6$, Eq. \eqref{eq1} then reads
\begin{eqnarray}
\label{eq9}
\mathcal{I}=\int\left[K\chi^{2}+\chi'^{2}
+\chi^{2}\alpha'^{2}+V(\chi)\right]\sqrt{-g}d^{4}x. 
\end{eqnarray}
Variation with respect to $\alpha$ and $\chi$, 
respectively, results in
\begin{eqnarray}
\label{eq10}
\alpha'&=&\frac{c_{\alpha}}{\chi^{2}}\\
\label{eq11}
2\chi''&=&2K\chi+2\alpha'^{2}\chi+\frac{dV}{d\chi},
\end{eqnarray}
where $c_{\alpha}$ is an arbitrary integration constant.
Employing Eq. \eqref{eq10} in \eqref{eq11}, multiplying by $\chi'$, 
and integrating we obtain the analog of the Friedmann 
equation
\begin{eqnarray}
\label{eq12}
\mathcal{Q}^{2}+\alpha'^{2}+K=\frac{V}{\chi^{2}}, 
\end{eqnarray}
where an integration constant has been absorbed 
in $V$. This contribution to $V$ corresponds to an 
EOS $w\equiv P/\rho=\frac{1}{3}$ 
as is evident from Eqs. \eqref{eq6} and \eqref{eq8}. 
Here, $P$ and $\rho$ are the pressure and energy density, 
respectively, and $\mathcal{Q}\equiv\frac{\chi'}{\chi}$ 
the analog of the conformal Hubble function, 
$\mathcal{H}\equiv\frac{a'}{a}$, has been defined. 
Making the replacement $\chi\rightarrow a$ and substituting 
$\alpha=0$ in Eqs. \eqref{eq9} and \eqref{eq12} we recover the 
EH action and standard Friedmann equation, respectively. 
Here, instead of space expansion, the entire cosmic evolution 
is due to the time-dependence of $\chi$, i.e. of active 
gravitational masses and the Planck mass, or equivalently 
due to the evolution of inertial masses, i.e. of the Rydberg 
`constant'. The evolution of the latter, on a static background, 
is then due to the monotonic growth of inertial masses in the 
redshifting era. Upon combining Eq. \eqref{eq12} and its derivative 
with Eq. \eqref{eq6} we obtain 
\begin{eqnarray}
\label{eq13}
2\mathcal{Q}'+\mathcal{Q}^{2}+K=
-\frac{3wV}{\chi^{2}}+3\alpha'^{2}. 
\end{eqnarray}

Back to the `Friedmann equations', 
Eqs. \eqref{eq12} and \eqref{eq13}. 
These are augmented with $\propto\alpha'^{2}$ terms 
that effectively behave as a negative energy with a 
`stiff' EOS ($w_{\alpha}=1$), with an effective 
contribution $-c_{\alpha}^{2}\chi^{-2}$ to the 
effective potential.    
Assuming this contribution is negligible 
at present, as well as at any observationally accessible 
cosmological era, then at sufficiently small $\chi$ this 
term competes with the radiation term, $V_{0}\propto\chi^{0}$, and at 
the `turning point' $\chi_{t}=c_{\alpha}/\sqrt{V}$ the rate 
$\mathcal{Q}$ momentarily vanishes and transition 
from $\mathcal{Q}<0$ to $\mathcal{Q}>0$ ensues.

Light element abundances set a limit on the redshift at the 
turning point, $z_{t}>10^{9}$. 
Since $c_{\alpha}^{2}/(\chi^{2}V_{0})\propto (1+z)^2$ 
it then follows that if $z_{t}>10^{9}$ then 
already by recombination $c_{\alpha}^{2}/\chi^{2}$ 
dropped at best to a trillionth of $V_{0}$ 
and therefore has virtually 
no impact on the standard $\Lambda$CDM 
description of structure formation history. 

Other bouncing models with similar behavior near the bounce, 
that contain a negative energy effective stiff matter 
component (that are sourced by other mechanisms), 
have been considered 
in \citeauthor{Barcelo and Visser2000} (\citeyear{Barcelo and Visser2000}) 
and \citeauthor{Barrow et al.2004} (\citeyear{Barrow et al.2004}).
We emphasize that in the present model the energy condition 
is not violated, i.e. $V$ -- the analog of $\mathcal{L}_{M}$ -- 
still includes only positive 
contributions, and it is only the contribution of 
the $U(1)$-symmetric kinetic term that makes an {\it effective} 
negative contribution, $-c_{\alpha}^{2}\chi^{-2}$, to the 
source term in Eq. \eqref{eq12}. This case is somewhat similar 
to the effective energy density associated with 
spatial curvature in the standard GR-based FRW spacetime, 
$\rho_{K}\equiv -K/H_{0}^{2}$; In case of 
a closed Universe, $K>0$, the effective $\rho_{K}$ is negative, 
and so a hypothetical radiation-only closed Universe 
(a legitimate solution of the Friedmann equation) 
will consequently undergo a bounce incurring no 
violation of the energy condition. Here, the negative 
kinetic energy associated with $\alpha$ is not counted 
as a contribution to the source, 
in the same fashion that $\mathcal{Q}^{2}$, 
the kinetic energy associated with $\chi$,  
(which in the standard Friedmann equation is 
represented by $a$) is not. 

Accounting for the vacuum-like, NR, radiation and 
effectively stiff energy densities, Eqs. \eqref{eq10} and \eqref{eq12} 
combine to 
\begin{eqnarray}
\label{eq14}
\mathcal{Q}^{2}=V_{4}\chi^{2}+V_{1}/\chi+V_{0}/\chi^2
-c_{\alpha}^{2}/\chi^{4}.
\end{eqnarray}
Two interesting limits of this equation 
will suffice for our purposes.  
The first is obtained by neglecting the 
vacuum-like and NR terms near the turning point. 
In this case, the Friedmann-like equation, Eq. \eqref{eq14}, 
integrates to
\begin{eqnarray}
\label{eq15}
\chi^{2}=V_{0}\eta^{2}+c_{\alpha}^{2}/V_{0},
\end{eqnarray}
where $\chi$ attains its minimum at $\eta=0$.
It could be readily integrated to yield the analog 
of the cosmic time around the turning point, 
$t\propto\int \chi(\eta)d\eta$, and is easily verified to be 
non-singular as well. In other words, the effective time 
coordinates of both massless and massive particles 
can be extended through the turning point, 
i.e. spacetime is geodesically-complete as is generically the 
case in non-singular bouncing cosmological models, 
\citeauthor{Ijjas and Steinhardt2018} (\citeyear{Ijjas and Steinhardt2018}). 
In the absence of turning point 
(e.g. in case that $\alpha$ is a fixed constant) 
the scalar field would have scaled $\chi\propto\eta$ 
as does the scale factor $a(\eta)$ in the radiation-dominated 
(RD) era. The scalar field then vanishes at $\eta=0$. 
Whereas this does not represent a curvature singularity, it is 
a topological singularity and it is not entirely clear that the theory 
can be extended past beyond $\eta=0$ as this 
would (at least formally) imply negative (active gravitational) masses, i.e. 
negative modulus of the scalar field, $\chi<0$. 

In the other extreme -- sufficiently far from the 
turning point, where the background dynamics is 
dominated by the quartic potential -- Eq. \eqref{eq12} 
integrates to 
$\chi=\left(\sqrt{\lambda}(\eta_{c}\mp\eta)\right)^{-1}$. 
The integration constant $\eta_{c}>0$ represents 
the start and end of the (conformal) time 
coordinate in the proposed nearly symmetric model, 
i.e. $\eta\in(-\eta_{c},\eta_{c})$. Again, since $\eta$ is 
bounded from below, this time due to the presence of the 
vacuum-like energy density component, 
then observed radial distances are bounded 
at $\eta=-\eta_{c}$ where the scalar field 
diverges and the model breaks down ($\eta=\mp\eta_{c}$ 
correspond to $t=\mp\infty$, i.e past and future infinity). 
Specifically, since in the case $K=0$ incoming null geodesics 
satisfy $r(\eta)=\eta_{0}-\eta$ and since $\eta>-\eta_{c}$ 
then the maximal observable distance 
at the present time is $r_{max}=\eta_{0}+\eta_{c}$. 
The latter can be much larger than the Hubble scale 
if $\eta_{c}\gg\eta_{0}$, thereby potentially addressing 
the horizon problem. 
In this regime $\mathcal{Q}=\frac{\dot{\chi}}{\chi}$ is constant and 
the scalar field is $\chi\propto\exp(\mathcal{Q}t)$. 
Recall that $\mathcal{Q}$ flips sign at the turning point 
and so $\chi$ diverges at $t\rightarrow\pm\infty$, exactly 
as does $a(t)$ in the standard FRW space-time 
in the DE-dominated phase.  
Although we focus here on DE-dominated asymptotics 
it is clear that a similar breakdown is 
expected in any model which is asymptotically 
dominated by $w<-1/3$ as Eq. \eqref{eq10} integrates in this case 
to $\chi=c\left(\eta_{c}\pm\eta\right)^{\frac{2}{1+3w}}$ 
(in the cases $\eta<0$ and $\eta>0$ respectively) 
where $c$ and $\eta_{c}$ are 
positive integration constants. Note that this is also the 
solution of the (equivalent) Friedmann equation, 
the scale factor $a(\eta)$, and this and 
scalar field singularity is present in $\Lambda$CDM 
as well (as expected; only the small $\chi$ 
dynamics is affected by introducing the phase $\alpha$). 
However, in the latter it is only a future singularity 
(where $\chi$ diverges), whereas 
in the proposed model it is also a past singularity taking 
place during the blueshifting epoch, thereby rendering 
the causally-connected Universe finite, although possibly 
much larger than naively thought based on 
the singular $\Lambda$CDM model. 

Regardless of whether the `flatness problem' is a genuine 
fine-tuning problem of the standard cosmological model or 
not, e.g. \citeauthor{Helbig2012} (\citeyear{Helbig2012}), 
\citeauthor{Carroll2014} (\citeyear{Carroll2014}),  
\citeauthor{Helbig2020} (\citeyear{Helbig2020}), 
we lay out the problem as it is normally presented in the literature 
followed by its resolution by the symmetric `bouncing' 
scenario in general, and the proposed model in particular. 
The `flatness problem' is often claimed to arise 
in the hot Big Bang model due to the monotonic expansion 
of space and the consequent 
faster dilution of the energy density of matter (either 
relativistic or NR) compared to the effective energy 
density dilution associated with curvature. It is thus hard 
to envisage how could space be nearly flat [as is indeed 
inferred from observations, e.g. 
\citeauthor{Planck Collaboration et al.2020} 
(\citeyear{Planck Collaboration et al.2020})] if not 
for an enormous fine-tuning taking place 
at the very early Universe, 
or alternatively for an early violent inflationary era. 
In the proposed scenario the matter content of the Universe 
has always been the same, and in particular the 
present ($\eta_{0}$) ratio of matter- to curvature-energy 
densities has been exactly the same at time $-\eta_{0}$. 
However, in the blueshifting era matter domination 
over curvature is actually an {\it attractor point} 
as the blueshifting universe starts essentially 
from $\chi\rightarrow\infty$. 
In other words, had the Universe been curvature-dominated (CD) 
at present (as is naively expected in the standard 
{\it expanding} hot Big Bang model in the absence of inflation), 
i.e. at $\chi_{0}$, it must have been CD 
at the `mirror time', $\eta=-\eta_{0}$, as well.
However, since $\rho_{M}\propto\chi$ 
while $\rho_{K}\propto\chi^{2}$ then a curvature 
domination at $-\eta_{0}$ would amount to an extremely 
fine-tuned $\rho_{M}/\rho_{K}\rightarrow 0$ at 
$\eta\rightarrow -\eta_{c}$, i.e. at the pre-bounce branch 
of $\chi\rightarrow\infty$.
As is well-known, entropy produced in the pre-bounce 
era could be processed at the bounce to thermal 
radiation, implying in effect 
that $\rho_{r}$ might somewhat 
change between pre- and post-bounce but the expectation 
in the proposed model is that the contracting and 
expanding epochs nearly mirror each other. 
This is especially the case if the bounce took 
place at sufficiently low temperature and density. 

We stress that not only does this {\it classical} `bouncing' 
cosmological model avoid 
the disturbing initial singularity problem, it also achieves this 
with no recourse to ideas from quantum gravity, a theory 
that does not currently exist (and in addition would probably 
require a hundred trillion times higher energy than is currently 
available in colliders to test). 

\subsection{Linear perturbation theory}
$\Lambda$CDM has successfully 
passed numerous tests and proved to be remarkably 
effective in explaining the formation and linear 
growth of density perturbations, predicting the CMB acoustic peaks, 
polarization spectrum, and damping features on small 
scales. It also correctly describes the linear and nonlinear 
evolution phases of the LSS (on sufficiently large scales) 
as well as the abundance of galaxy and galaxy cluster halos. 
Therefore, it would seem essential to establish 
equivalence of linear perturbation theory between 
the model proposed here and $\Lambda$CDM. 

Consider linear perturbations over the 
FRW spacetime in the comoving frame.
Metric perturbation variables include the scalars 
$\varphi$ and $\psi$, vector mode 
$v_{i}$ and tensor modes $h_{ij}$, where the latter 
are subject to the constraint $\gamma^{ij}h_{ij}=0$.
The weakly perturbed line element is 
$ds^{2}=-(1+2\varphi)d\eta^{2}+2v_{i}d\eta dx^{i}
+[(1-2\psi)\gamma_{ij}+2h_{ij}]dx^{i}dx^{j}$, 
with $\gamma_{ij}\equiv diag[1/(1-Kr^{2}),r^{2},
r^{2}\sin^{2}\theta]$. When stress is negligible 
then $\varphi=\psi$.
We define the fractional energy density and 
pressure perturbations (in energy density units) 
$\delta_{\rho_{M}}\equiv\delta\rho_{M}/\rho_{M}$ and 
$\delta_{P_{M}}\equiv\delta P_{M}/\rho_{M}$ ($=w\delta_{\rho_{M}}$), 
respectively. The matter velocity is $v$. 

Transforming from the frame where $G=constant$ [in which case 
Eq. \eqref{eq1} reduces to GR] to an arbitrary field frame with 
$\Omega(x)=1-\delta_{\chi}(x)$, and assuming 
$\delta_{\chi}\equiv\frac{\delta\chi}{\chi}\ll 1$, 
implies in particular that the scalar and metric field  
transform as $\chi\rightarrow\chi/\Omega\approx\chi(1+\delta_{\chi})$ 
and $g_{\mu\nu}\rightarrow\Omega^{2}g_{\mu\nu}$, respectively, 
i.e. $\psi\rightarrow\psi+\delta_{\chi}$, and 
$\delta_{\rho_{M}}\rightarrow\delta_{\rho_{M}}+4\delta_{\chi}$ 
to leading order. 
Consequently, the new, `shifted', perturbation variables, 
e.g. $\tilde{\varphi}\equiv\varphi+\delta_{\chi}$, 
$\tilde{\psi}\equiv\psi+\delta_{\chi}$, 
$\tilde{\rho}_{M}\equiv\delta_{\rho_{M}}+4\delta_{\chi}$
etc., obey the same perturbation equations 
that are satisfied by the old perturbation quantities.
The fact that the structure of the perturbation 
equations is unchanged under Weyl transformations 
is crucial in the context of stability near the 
bounce because perturbations of the scalar 
field in bouncing scenarios based on scalar-tensor 
theories of gravity are a potential cause for 
instability near the bounce, 
e.g. \citeauthor{Gratton et al.2004} (\citeyear{Gratton et al.2004}). 

By virtue of the U(1) symmetry we have two new 
perturbation variables $\delta_{\chi}$ and $\delta\alpha$ 
in addition to the standard perturbation 
variables $\varphi$, $\psi$, $\delta\rho_{M}$ 
and $v$. However, in the long wavelength limit, 
and ignoring coupling with other perturbation variables, 
Eq. \eqref{eq10} $\alpha'=c_{\alpha}/\chi^{2}$ implies that 
$\frac{\delta\alpha'}{\alpha'}
=-2\delta_{\chi}$, and consequently the only 
new independent perturbation variable is, 
e.g. $\delta_{\chi}$ in this limit. 
Possible implications of redefining the gravitational 
potential to CDM on galactic and cosmological scales in this 
framework are discussed in 
\citeauthor{Shimon2021a} (\citeyear{Shimon2021a}) 
and \citeauthor{Shimon2022a} (\citeyear{Shimon2022a}).
Here we point out 
that the same rational can be applied on cosmological 
scales to the point that particle CDM may not be required 
at the perturbation level for explaining the clustering 
and growth of structure because $\delta_{\chi}$ is {\it a priori} 
an arbitrary function of space and time. 
Specifically, the spectrum 
of metric perturbations at recombination, with an rms 
level of $O(10^{-5})$, may simply reflect a similar 
perturbation in $G$, gauged by $\delta\chi$, 
rather than the existence of particulate CDM. 
And indeed, $G$ has never been {\it directly} 
measured on cosmological scales, 
surely not at the $O(10^{-5})$ precision, and the fact 
that CDM seems at all to be required on these scales 
heavily relies on the premise that $G$ is a Universal constant. 
All this implies that on the largest scales, 
and ignoring coupling between the various perturbation 
variables at the leading perturbation order 
both $\varphi$ and 
$\frac{\delta\alpha'}{\alpha'}$ are perturbed at 
the $O(10^{-5})$ level at the turning point and consequently the 
perturbations are stable. A more detailed stability analysis 
is carried out in Appendix A.

It has been argued that, quite generally, 
modified theories of gravity that include no CDM 
component ought to be unnaturally 
contrived, exhibiting strong features in the 
transfer function \citeauthor{Pardo and Spergel2020} 
(\citeyear{Pardo and Spergel2020}). 
In arriving at this conclusion it has been assumed that 
the energy-momentum is conserved 
and that acceleration is caused by baryon-matter-only 
overdensities. However, none of these assumptions apply 
to the framework proposed in the present work.

In any case, we emphasize that the model proposed here by 
no means relies on the premise that CDM phenomena are merely 
a manifestation of variations of $\chi$. CDM can be equally 
well taken to be particulate, exactly as in $\Lambda$CDM 
[Eq. \eqref{eq12} is agnostic to the microphysics of 
CDM insofar it is characterized by $w=0$].
Rather, it is only mentioned in passing that at both 
the background and perturbation levels CDM phenomena can be accounted for by 
$\chi$ and its perturbations. 

In $\Lambda$CDM the linear gravitational 
potential has two modes in the matter-dominated (MD) era, $\varphi=constant$ 
and $\varphi\propto\eta^{-5}$. The latter dies off very quickly 
in an expanding Universe and is therefore conventionally 
ignored in practice. However, in a contracting Universe 
$\varphi\propto\eta^{-5}$ is the fastest growing mode, and if we assume that 
primordial scalar perturbations are set sometime during the 
contracting MD phase or even at $-\eta_{c}$, deep in 
the DE era of the contracting phase, they grow by a 
factor $(1+z_{eq})^{5/2}$ during the MD era 
(and essentially freeze during the DE and RD 
pre-bounce epochs). Assuming the model is 
approximately symmetric around the bounce 
and employing the observationally favored value 
for the redshift at radiation-matter equality, 
$z_{eq}\approx 3400$, we obtain that $\varphi$ 
amplifies at most by a factor billion over 
the time period spanned between $-\eta_{0}$ 
and $-\eta_{eq}$. Consequently, scalar perturbations 
which started anywhere in the range $\varphi\in(10^{-14}-10^{-5})$ 
(depending on their formation epoch)
would end up at the observed $\varphi=O(10^{-5})$ level 
by the time of recombination, $\eta_{rec}$.

\section{SUMMARY}

While $\Lambda$CDM has clearly been very successful in 
phenomenologically interpreting 
a wide range of observations, it still lacks a 
microphysical explanation of 
several key components, primarily the nature of CDM and DE.
It also suffers from a few outstanding conceptual 
problems such as the initial singularity, in addition 
to a few `coincidence' or `naturality' problems.

{\it Direct} spectral information on the CMB is 
unavailable (due to opacity) in the pre-recombination 
era ($z\gtrsim 1100$). 
From the observed cosmic abundance of light elements, 
BBN at redshifts $O(10^{9})$ could be indirectly probed. 
Earlier on, at $z=O(10^{12})$ 
and $z=O(10^{15})$ [energy scales of $O(200)$ MeV 
and $O(100)$ GeV, respectively], 
the quantum chromodynamics (QCD) and electroweak 
phase transitions 
had presumably took place, although their (indeed weak) 
signatures in, e.g. the CMB, have not been found. 

In addition, inflation, a cornerstone of the standard 
cosmological model, is clearly beyond the realm of 
well-established physics; its ultimate detection via the B-mode 
polarization that it imprints on the CMB could be 
achieved only if inflation took place at energy scales 
$\sim$trillion times larger than currently 
achievable in colliders. Moreover, theoretical 
expectations for the amplitude of this B-mode as 
a function of the energy scale of inflation rely 
on the assumption that gravitation is genuinely 
quantized. The latter assumption lacks empirical 
basis at present. By itself, inflation is plagued  
by the the $\eta$-problem, trans-Planckian problem, and 
the `measure problem' in the multiverse, essentially 
a lack of predictive power. 

Ideally, an alternative cosmological model that 
agrees well with $\Lambda$CDM at BBN energies 
and lower, i.e. $z<10^{10}$, while 
still addressing the classical problems of the hot 
Big Bang model that inflation was originally 
designed to undertake, 
as well as avoiding the initial curvature singularity, 
and all this in the $\lesssim 10TeV$ range of energies, 
will be an appealing alternative 
that relies on experimentally established physics. 
One conclusion of the present work is that this could 
be in principle achieved, at least in part, with a (relatively 
late {\it classical}) non-singular `bounce' that also removes 
the technically and conceptually undesirable initial 
singularity problem of GR-based cosmological 
models. In order to achieve such a bounce 
within GR, or a conformally-related theory, certain 
`energy conditions' have to be effectively violated. 
One specific realization of this program has been 
the focus of the present work.

Symmetries play a key role in our current understanding 
of the inner workings of the fundamental 
interactions. For example, the SM of particle physics is 
based on a {\it local} $U(1)\times SU(2)\times SU(3)$ 
gauge group with quantized gauge fields. In addition, 
our favorite theory of gravitation, GR, is 
diffeomorphism-invariant. In this work we entertained 
the possibility that in addition to diffeomorphism-invariance, 
GR is endowed with Weyl invariance, i.e. {\it local} scale 
invariance, as well as an internal $U(1)$ symmetry with a global 
charge. 
The SM of particle physics is assumed as is, with an explicitly 
broken Weyl symmetry.
Only the salient merits of the cosmological model based on this 
alternative theory of gravitation have been discussed in 
the present work. 

In the proposed model spacetime is described by the FRW 
metric in comoving frame, which in the absence of spatial 
curvature reduces to the Minkowski spacetime. Here, 
the role of the scale factor in $\Lambda$CDM as the regulator 
of cosmic history is played by 
the modulus, $\chi$, of a complex scalar field 
$\phi$ that lives on a static background. 
The phase, $\alpha$, plays a crucial role 
near the turning point and is largely irrelevant elsewhere.
There is no analog to $\alpha$ in $\Lambda$CDM. 
Here, $\chi$, which regulates the evolution 
of (dynamical) active gravitational masses starts 
infinitely large, first monotonically decreases until 
it `bounces', then grows again without bound. 
Put alternatively, the Planck length starts out 
infinitely small, increases until it peaks at 
the turning point, then decreases again. 
Described in terms of these length `units' the Universe is 
said to undergo a `contraction' phase of evolution, 
followed by a `bounce' and `expansion'.
Since spacetime is described by the FRW metric in the comoving 
frame, i.e. the time coordinate is conformal, then inertial 
masses evolve exactly as do active gravitational masses and the 
Planck mass. Cosmological redshift is then a manifestation of 
evolving Rydberg `constant' rather than space expansion.

The alternative cosmological scenario explored 
in this work starts with a deflationary evolution which 
culminates in a turning point 
when the (absolute value of the negative) energy 
density associated with the effective `stiff matter' 
(provided by the kinetic term of $\alpha$) 
momentarily equals that of radiation. 
In the vacuum-like-dominated epoch the energy density 
of the Universe is dominated by a $\propto\chi^{4}$ term 
in the potential which is genuinely classical with no quantum 
fluctuations. Therefore, DE according to the present scenario 
is not zero-point energy but rather a manifestation of the 
self-coupling of the scalar field, i.e. a term in the potential of the 
form $V_{4}\chi^{4}$ with $V_{4}$ being a dimensionless parameter.
This DE contribution is characterized by a non-dynamical 
EOS with no recourse to, e.g. 
a new quintessence, field; here, the same scalar field accounts 
for both `$G$' and DE, and possibly also CDM. 

One of the most notable achievements of inflation was 
the realization that a slightly red-tilted primordial 
spectrum of gaussian perturbations can be generated by quantum 
fluctuations in a vacuum-like expanding Universe.
It should be note that a decade before the inflationary 
scenario has been proposed it has been realized by Harrison and 
Zeldovich, as well as by others, that the primordial spectrum of 
perturbations should be nearly flat for the rms perturbations 
on the wide dynamical range of cosmological scales not to have 
infrared or ultraviolet divergences that mark 
the breakdown of linear perturbation theory.
In any case, no mechanism has been proposed in the present 
work for the generation of primordial density perturbations.
Whatever this mechanism turns out to be, 
it does not involve quantum fluctuations 
of the metric field, unlike in inflation. 
Again, it is assumed here that gravitation 
is a genuine classical interaction. 

In addition, the `anisotropy problem' that in general plagues 
bouncing scenarios does not exist in our construction. 
As discussed above, Weyl symmetry and the consequent 
absence of any dimensional parameter in the action, 
in addition to the postulated global U(1) symmetry, protects 
the model from running into a chaotic, anisotropy-dominated, 
evolution phase, unless we are willing to consider 
non-canonical terms in the potential, $V$, of the form, e.g. 
$V_{ani}\propto(\bar{\psi}\psi)^{2}\chi^{-2}$ and 
$C_{\alpha\beta}^{\ \ \gamma\delta}C_{\gamma\delta}^{\ \ \rho\sigma}
C_{\rho\sigma}^{\ \ \alpha\beta}\chi^{-2}$, where 
$\psi$ is e.g., a Dirac field, and $C_{\rho\sigma}^{\ \ \alpha\beta}$ 
is the Weyl tensor.

Conformal time is both past- and future-bounded in 
this scenario, i.e. $\eta\in(-\eta_{c},\eta_{c})$.
In principle, any `horizon problem' could be avoided 
if $\eta_{c}\gg \eta_{0}$. 
Specifically in this scenario, cosmic history starts with very large 
(and in principle infinite) particle masses and therefore 
the causal horizon is much larger than would be 
naively expected from monotonically 
growing masses (that corresponds to the redshifting era), 
i.e. essentially $\eta_{0}\ll\eta_{c}+\eta_{0}$ 
if $\eta_{c}\gg\eta_{0}$. 
Likewise, the `flatness problem' afflicting 
the hot Big Bang scenario stems from the 
slower decay of the energy density 
associated with curvature as compared to 
that of matter in a {\it monotonically} expanding Universe. 
In bouncing scenarios the 
situation is reversed before the turning point; 
starting at infinitely large $\chi$ 
(masses) one typically expects to 
find that the energy density in the forms of NR matter and radiation  
largely exceeds that of curvature at any {\it finite} 
$\chi$ value in the blueshifting era. 
Since this adiabatic model is nearly symmetric 
in $\chi$ around the turning point (barring possible entropy 
processing effects at around the RD era, the efficacy of which 
depends on the redshift at which such a bounce takes place), 
one generally expects the Universe to 
look spatially flat at any finite $\chi$ after the would-be 
singularity (actually a non-singular `bounce'). 
From this perspective flatness is an 
attractor-, rather than an unstable-point that 
requires fine-tuning of the initial conditions. 

The model we considered is falsifiable in several ways: 
First, $w_{DE}=-1$ due to Weyl invariance and any 
observationally inferred $w_{DE}\neq -1$ would either 
rule out this particular model or alternatively either 
imply soft breakdown of WI at very early and late times or 
the existence of a non-canonical DE term in the potential of the form, 
e.g. $V_{DE}\propto(\bar{\psi}\psi)^{-\varepsilon/3}
\chi^{4+\varepsilon}$ (where $\varepsilon$ is some 
dimensionless parameter), that involves non-integer, 
and possibly irrational, powers of the fields. 
Second, within the framework adopted here, 
if a scale-invariant B-mode 
polarization is ultimately 
measured, it would provide a compelling evidence that 
gravity is quantized, in contradiction to the assumption 
made here that gravity is a genuinely classical interaction 
which implies that its perturbations are not subject to 
the Bunch-Davies vacuum condition. Consequently, unlike 
the inflationary-induced 
B-mode polarization of $\Lambda$CDM, it does not follow 
from any fundamental principle that B-mode polarization 
have to be characterized by a flat spectrum.
Third, signals from primordial phase transitions as well 
as leptogenesis or baryogenesis that ought to be 
imprinted in the CMB anisotropy and polarization 
(perhaps too weak to be detected) may not have taken 
place at all in the proposed model, depending on 
the typical temperature at the turning point.

We believe that, in addition to addressing the 
cosmological horizon and flatness problems, the 
framework proposed here provides 
important insight on the nature of DE, 
and initial singularity and stability near 
the turning point. Even so, the work presented here is 
by no means exhaustive, and indeed a few of its basic aspects 
will be further elucidated in future works. 

%\section*{Acknowledgments}
%The author is indebted to Yoel Rephaeli for 
%numerous constructive, critical, and 
%thought-provoking discussions which were invaluable for this work.
%This work has been supported by the Joan and Irwin 
%Jacobs donor-advised fund at the
%JCF (San Diego, CA).

\section*{APPENDIX A: STABILITY ANALYSIS}

As discussed in section 3.2 the initial scalar metric
perturbations deep in the pre-bounce era should be 
at the $O(10^{-14})-O(10^{-5})$ level (depending 
on their formation time in the blueshifting phase) 
so that by the time of recombination they have grown 
to $O(10^{-5})$. 
In this Appendix we analyze the dynamics of scalar 
perturbations at and near $\eta=0$ 
and show that all linear perturbation variables smoothly 
transform through this point.

Near this point the energy budget is 
dominated by radiation and the kinetic 
term associated with $\alpha$. 
The Friedmann-like equation, Eq. \eqref{eq12}, then integrates to
\begin{eqnarray}
\label{eq16}
\chi^{2}=A\eta^{2}+B, 
\end{eqnarray}
where according to Eq. \eqref{eq15} $A\equiv V_{0}$ and 
$B\equiv c_{\alpha}^{2}/V_{0}$.
The analog of conformal Hubble function in this case 
becomes $\mathcal{Q}=A\eta/(A\eta^{2}+B)$.
Linear perturbation equations [e.g., \citeauthor{Hwang and Noh2001} 
(\citeyear{Hwang and Noh2001})] result in coupled 
equations for $\delta\alpha$ and $\varphi$
\begin{eqnarray}
\label{eq17}
\delta\alpha''+\frac{2A\eta}{A\eta^{2}+B}\delta\alpha'+k^{2}\delta\alpha
&=&-\frac{4\sqrt{AB}}{A\eta^{2}+B}\varphi'\\
\label{eq18}
\varphi''+\frac{4A\eta}{A\eta^{2}+B}\varphi'+\frac{k^{2}\varphi}{3}
&=&\frac{2\sqrt{AB}}{A\eta^{2}+B}\delta\alpha',
\end{eqnarray}
where it is understood that fractional perturbations of the modulus 
scalar field, $\delta_{\chi}$, are absorbed in the gravitational 
potential as described in section 3.2.
Eq. \eqref{eq18} is a generalized Bardeen equation that is obtained 
from combination of the Arnowitt-Deser-Misner (ADM) 
energy constraint and the Raychaudhuri equation.
In the long wavelength limit this system of equations 
is satisfied by
\begin{eqnarray}
\label{eq19}
\varphi&=&c_{1}+\frac{c_{2}\eta}{(B+A\eta^{2})^{2}}
+\frac{c_{3}(A\eta^{2}-B)}{(B+A\eta^{2})^{2}}\nonumber\\
\frac{\delta\alpha'}{\alpha'}&=&-\frac{4c_{2}\eta}{(B+A\eta^{2})^{2}}
-\frac{c_{3}(-3B^{2}+6AB\eta^{2}+A^{2}\eta^{4})}{B(B+A\eta^{2})^{2}},
\end{eqnarray}
where $c_{1}$, $c_{2}$ and $c_{3}$ are three integration constants. 
It is clear from Eqs. \eqref{eq19} that both $\varphi$ 
and $\delta\alpha'/\alpha'$ are well-behaved at $\eta=0$ due 
to the non-vanishing $B$, i.e. the non-vanishing of $\alpha'$, 
and perturbation theory does not break down there. 

To solve for the short-wavelength limit we differentiate Eq. \eqref{eq17} 
with respect to $\eta$ and then substitute for $\delta\alpha'$ from Eq. \eqref{eq18}. The resulting fourth-order equation for $\varphi$ is
\begin{eqnarray}
\label{eq20}
&&\left(\frac{Ax^{2}}{k^{2}}+B\right)\varphi+\frac{18A}{k^{2}}x\varphi_{x}
+\left(\frac{60A}{k^{2}}+\frac{4A}{k^{4}}x^{2}+4B\right)\varphi_{xx}\nonumber\\
&+&\frac{30A}{k^{2}}x\varphi_{xxx}+3\left(B+\frac{Ax^{2}}{k^{2}}\right)\varphi_{xxxx}=0, 
\end{eqnarray}
where e.g., $\varphi_{x}\equiv\frac{\partial\varphi}{\partial x}$, 
in the large-$k$ limit, which has no closed-form solution. Here $x\equiv k\eta$.
In the limit $x\ll 1$, i.e. $\eta\rightarrow 0$, the potential $\varphi$ 
decouples from both $A$ and $B$, 
and the equation considerably simplifies to
\begin{eqnarray}
\label{eq21}
3\varphi_{xxxx}+4\varphi_{xx}+\varphi=0, 
\end{eqnarray}
with the general solution
\begin{eqnarray}
\label{eq22}
\varphi=c_{1}\cos x+c_{2}\sin x+c_{3}\cos\left(x/\sqrt{3}\right)
+c_{4}\sin\left(x/\sqrt{3}\right), 
\end{eqnarray}
illustrating the fact that perturbations are manifestly 
stable on small scales as well.

Slightly off $\eta=0$, $A\eta^{2}\gtrsim B$ perturbations 
smoothly approach their short wavelength limit standard 
behavior in the RD era. In the latter, within $\Lambda$CDM, 
$\varphi=\frac{1}{x}\left[c_{j}j_{1}(x)+c_{y}y_{1}(x)\right]$ 
where $j_{1}$ and $y_{1}$ are the spherical Bessel and the 
modified spherical Bessel functions, respectively, of the first kind.
In $\Lambda$CDM, $c_{y}=0$ since $y_{1}(0)$ diverges, 
and the phase of the oscillating potential is 0, 
i.e. $\varphi\propto\sin x$. This needs not 
be the case in the proposed model as is evident from this discussion 
insofar $B\neq 0$. The implications for CMB observables of not 
neglecting the `diverging' mode have been discussed recently by 
\citeauthor{Kodwani et al.2019} (\citeyear{Kodwani et al.2019}).

\label{lastpage}
\end{document}